\documentclass[journal = ancac3]{article}
\usepackage[etalmode=truncate,maxauthors=20]{achemso}
\setkeys{acs}{articletitle = true}
\usepackage[version=3]{mhchem}
\usepackage[left=1.5cm, right=1.5cm, top=1.785cm, bottom=2.0cm]{geometry}
\usepackage{mathptmx}
\usepackage{graphicx} 
\usepackage{float}
\usepackage[english]{babel}
\usepackage{charter}
\usepackage[compact]{titlesec}
\usepackage{hyperref}

\begin{document}
\begin{center}
\LARGE{\textbf{Reversible Formation of Thermoresponsive Binary Particle Gels with Tunable Structural and Mechanical Properties}}
\end{center}

\makeatletter
\def\nlfootnote{\xdef\@thefnmark{}\@footnotetext}
\makeatother

\begin{center}
\large{Jasper N. Immink$^{\dagger}$, J. J. Erik Maris$^\ddagger$, J\'{e}r\^{o}me J. Crassous$^{\perp}$, Joakim Stenhammar$^\dagger$, and Peter Schurtenberger$^{\dagger,\parallel}$}\\
$\dagger$ \normalsize{\textit{Division of Physical Chemistry, Lund University, SE-22100, Lund, Sweden.}}\\
$\ddagger$ \normalsize{\textit{Inorganic Chemistry and Catalysis, Utrecht University, 3584CG, Utrecht, the Netherlands.}}\\
$\perp$ \normalsize{\textit{Institute of Physical Chemistry, RWTH Aachen University, 52074, Aachen, Germany.}}\\
$\parallel$ \normalsize{\textit{Lund Institute of advanced Neutron and X-ray Science (LINXS), Lund University, SE-22100, Lund, Sweden}}

\end{center}

\section*{Abstract}
We investigate the collective behavior of suspended thermoresponsive microgels, that expel solvent and subsequently decrease in size upon heating. Using a binary mixture of differently thermoresponsive microgels, we demonstrate how  distinctly different gel structures form, depending on the heating profile used. Confocal laser scanning microscopy (CLSM) imaging shows that slow heating ramps yield a core-shell network through sequential gelation, while fast heating ramps yield a random binary network through homo-gelation. Here, secondary particles are shown to aggregate in a monolayer fashion upon the first gel, which can be qualitatively reproduced through Brownian dynamics simulations using a model based on a temperature-dependent interaction potential incorporating steric repulsion and van der Waals attraction. Through oscillatory rheology it is shown that secondary microgel deposition enhances the structural integrity of the previously formed single species gel, and the final structure exhibits higher elastic and loss moduli than its compositionally identical homo-gelled counterpart. Furthermore, we demonstrate that aging processes in the scaffold before secondary microgel deposition govern the final structural properties of the bigel, which allows a detailed control over these properties. Our results thus demonstrate how the temperature profile can be used to finely control the structural and mechanical properties of these highly tunable materials.\\

\noindent \textit{Keywords: microgel - binary particle gel - responsive gel - gel microstructure - sequential particle gelation}\\

\noindent Materials with tunable or switchable properties have intrigued material scientists and industrial engineers for decades. Colloidal suspensions are a precursor for such switchable systems, allowing formation of colloidal crystals, glasses or gels. All of these systems exhibit strongly varying collective properties such as elasticity, viscosity or optical properties.\cite{Aksay1990,vanBlaaderen2002} One strategy to introduce \textit{in-situ} external tunability is using poly-N-isopropylacrylamide (pNIPAm) microgel particles, which are cross-linked polymer networks of colloidal size. pNIPAm forms hydrogen bonds with water at room temperature, causing the microgel particles to swell.\cite{Pelton1986} Upon heating above the volume phase transition temperature ($T_{\text{VPT}}$) of 32~$^\circ$C, the hydrogen bonds are thermally disrupted, the microgel expels water and significantly reduces in size.\cite{Shirahama1993} At high ionic strengths, the temperature-induced collapse is accompanied by a reversible transition from soft repulsive to attractive interactions.\cite{Hu2003} 
Controlled destabilization upon heating thus enables the reversible formation of elastic, load-bearing colloidal gel networks that redisperse and lose elasticity upon cooling and reswelling.\cite{Richtering2000,Weitz2010}

Despite decades of research into gelating colloids, a full understanding of the kinetics of these systems has proven elusive.\cite{Schmiedeberg2016} Research points towards a gel formation that follows a path of spinodal decomposition and subsequent arrest,\cite{Giglio1992, Schurtenberger2007} while the formation kinetics and final structure have proven hard to predict.\cite{Zukoski2003,Mohraz2016} Similar arrest kinetics are also observed in glasses, which show similarities to gels in structure and dynamics\cite{Weitz2005} and the subject of caging effects is relevant to understanding both glassy and gel properties. Here, binary colloidal mixtures have seen significant interest: in asymmetric hard sphere systems, the smaller particle can act as lubricant and the mobility of a particle is affected by the local cage and its constituents,\cite{Weeks2011} while in soft colloidal mixtures, new cage organizations as compared to hard sphere systems were found, where osmotic pressure of smaller particles causes the cages to become anisotropic.\cite{Vlassopoulos2008,Cloitre2014} Furthermore, varying the components in asymmetrically soft colloidal mixtures can yield re-entrant glassy dynamics and arrested phase separation, both topics of considerable interest.\cite{Vlassopoulos2018} Apart from the complex arrest kinetics, macroscopic gel formation is often an irreversible process\cite{Zaccarelli2007,Tartaglia1995} and several open questions remain in terms of macroscopic gel structure and rheological properties,\cite{Morbidelli2004} especially at high densities.\cite{Scheffold2014}

In this study, we report results from an investigation of a binary microgel suspension that allows for the reversible formation of core-shell colloidal gel structures. The microgels used consist of either cross-linked pNIPAm polymers or poly-N-isopropylmethacrylamide (pNIPMAm) polymers, the latter of which has been shown to form microgels behaving similarly to pNIPAm microgels, but with a slightly elevated $T_{\text{VPT}}$ of approximately 45~$^\circ$C.\cite{Pichot1999} We build upon the existing knowledge of sequentially gelating colloidal bigels using temperature-controlled deposition\cite{Sprakel2015,Urayama2018} and DNA-hybridization\cite{Foffi2012, Foffi2014, Eiser2013} and show how the details of the temperature profile can be used to control the structural and mechanical properties of these bigels. We demonstrate this using a combination of confocal laser scanning microscopy (CLSM), Brownian dynamics simulations and oscillatory rheology. Using slow temperature ramping, our gels are formed through sequential aggregation and conform to the structure of the earlier formed single particle gel, enhancing its elastic properties and stability and allowing fine control over mechanical properties. The reversible nature and facile manipulation of these materials holds great promise for the design of tunable colloidal materials and the understanding of gel formation and aging kinetics.

\section*{Results and discussion}
\begin{table}
\centering
\caption{Particles used in experiments, with their polymer, dye type, and hydrodynamic diameters in collapsed and swollen state according to dynamic light scattering at low ionic strengths of 10$^{-3}$M KCl in order to prevent aggregation in the collapsed state. Swelling curves can be found in the Supporting Information.$^\dag$}
\begin{tabular}{ l  l  l  l  l  l  l }
\hline
Polymer & Abbreviation & Dye & $\sigma_H$(20~$^\circ$C) [nm] & $\sigma_H$(50~$^\circ$C) [nm]\\ 
\hline
pNIPAm & pN & Rhodamine & 620 & 354\\										
pNIPMAm & pM & Fluorescein & 573 & 345\\										
\end{tabular}
\label{Methods-parttable}
\end{table}

\subsection*{Confocal Laser Scanning Microscopy}
\begin{figure}
	\centering
	\includegraphics{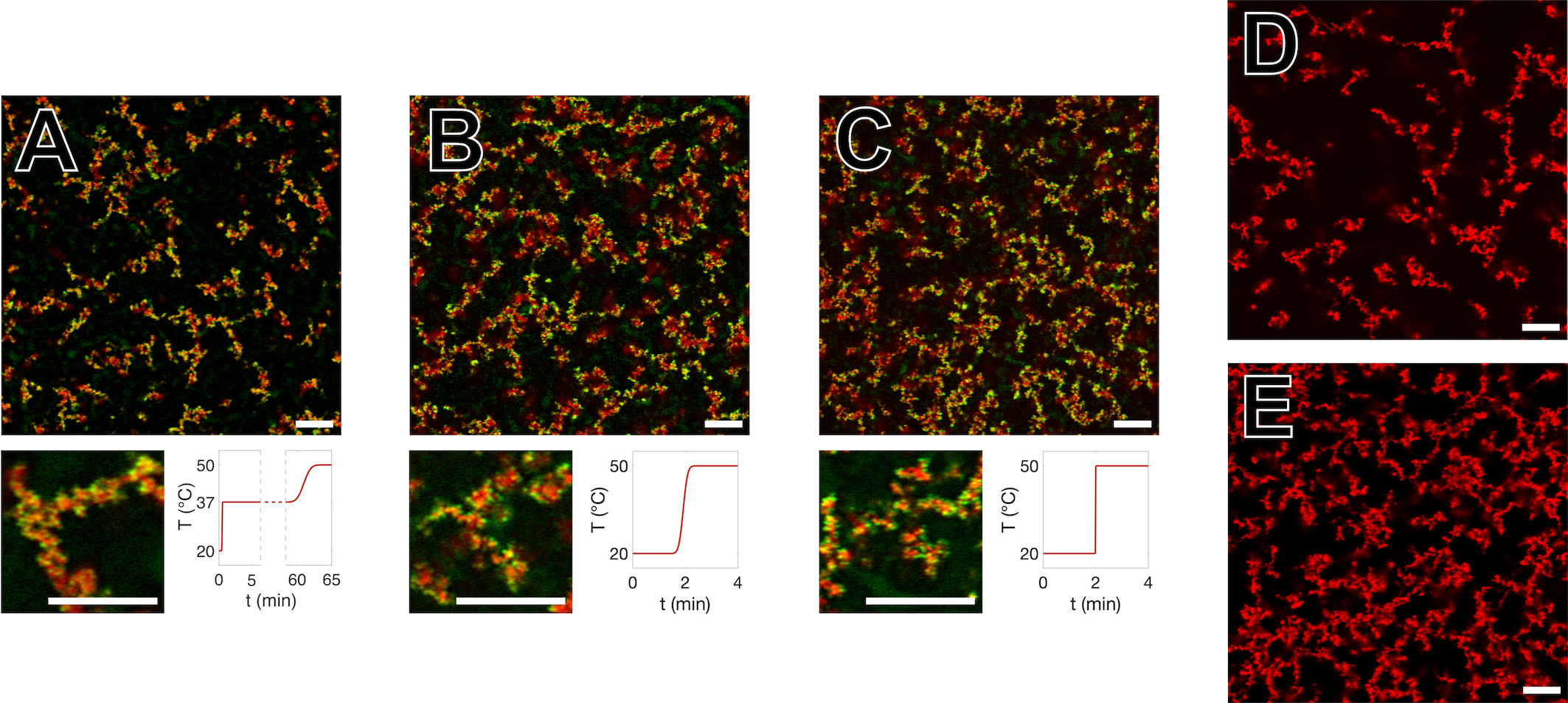}
	\caption{Confocal images of five microgel suspensions, heated using different temperature profiles. (A): An overview and a detail of a binary sample with $\phi_{50}(\text{pN}) = \phi_{50}(\text{pM}) = 0.025$ that was heated slowly, resting at 37~$^\circ$C for 60 minutes, according to the accompanying temperature profile. pN particles have had the time to form a separate gel structure before pM particles collapse. (B): A compositionally identical sample to (A) which has instead been heated over the course of two minutes, similar to the fastest quench possible in rheology, as described later. (C): Another compositionally identical sample to (A) and (B) that has been heated instantaneously. (D): A single pN particle gel at $\phi_{50}(\text{pN}) = 0.025$ shows a comparable structure to the backbone in (A). (E): A single pN particle gel at $\phi_{50}(\text{pN}) = 0.05$, with a structure similar to the sample in image (C). All scale bars correspond to 10 $\mu$m and zoom-ins are not necessarily part of the larger corresponding picture.}
	\label{RES-confocalpic}
\end{figure}
In order to prepare gel structures for CLSM imaging, two types of microgels were prepared according to the procedure in the Methods section and their properties can be found in Table~\ref{Methods-parttable}. Binary microgel dispersions of pNIPAm and pNIPMAm microgels (from now on abbreviated as pN and pM) in strongly screened conditions (10$^{-1}$M KCl) were heated from 20~$^\circ$C to 50~$^\circ$C using slow or fast temperature profiles. A binary sample of $\phi_{50}(\text{pN}) = \phi_{50}(\text{pM}) = 0.025$, with $\phi_{50} \equiv $ the volume fraction at $T = 50 ^\circ$C, that was heated by halting the quench for 60 min at 37~$^\circ$C before further heating to 50~$^\circ$C yielded CLSM images of a core-shell bigel network with the red pN particles in the center, decorated by a monolayer of the pM particles (Fig.~\ref{RES-confocalpic}A), in accordance with previous observations.\cite{Sprakel2015, Eiser2013} Fig.~\ref{RES-confocalpic}B represents a quench that occurs over the course of two minutes, and shows a decrease in mesh size compared to (A). Finally, Fig.~\ref{RES-confocalpic}C represents a sample that was brought from 20~$^\circ$C to 50~$^\circ$C instantaneously. This sample shows a further decrease in mesh size, and close inspection shows clusters of pN particles covered by pM particles. Due to the short time scale of this quench, pN particles aggregate and cluster, but do not have time to form a continuous scaffold before the pM particles start associating in bulk. The full reversibility of the system allows for switching between the core-shell structure and the more randomized network by cooling, subsequent redispersion, and heating using a different temperature profile. 
\\
As noted, it is apparent from Figs.~\ref{RES-confocalpic}A and \ref{RES-confocalpic}C that not only the local structure but also the larger backbone structure is strongly affected by the heating profile. The slow quench backbone structure (Fig.~\ref{RES-confocalpic}A) is similar to backbone of a pure pN particle gel of $\phi_{50}(\text{pN}) = 0.025$ (Fig.~\ref{RES-confocalpic}D), and the fast quench backbone (Fig.~\ref{RES-confocalpic}C) is similar to a pure $\text{pN}$ gel of $\phi_{50}(\text{pN}) = 0.05$, (Fig.~\ref{RES-confocalpic}E). This indicates that the slow quench gel is formed by decorating the already present pN particle gel, whereas the fast quench gel is formed through a homogeneous gelation, comparable to the gel observed at $\phi_{50}(pN) = 0.05$. This leads to thicker but longer and sparser strands for the slow quench, and thinner but shorter and more numerous strands for the fast quench gels. 

\subsection*{Model and simulation}
\begin{figure}
\centering
\includegraphics{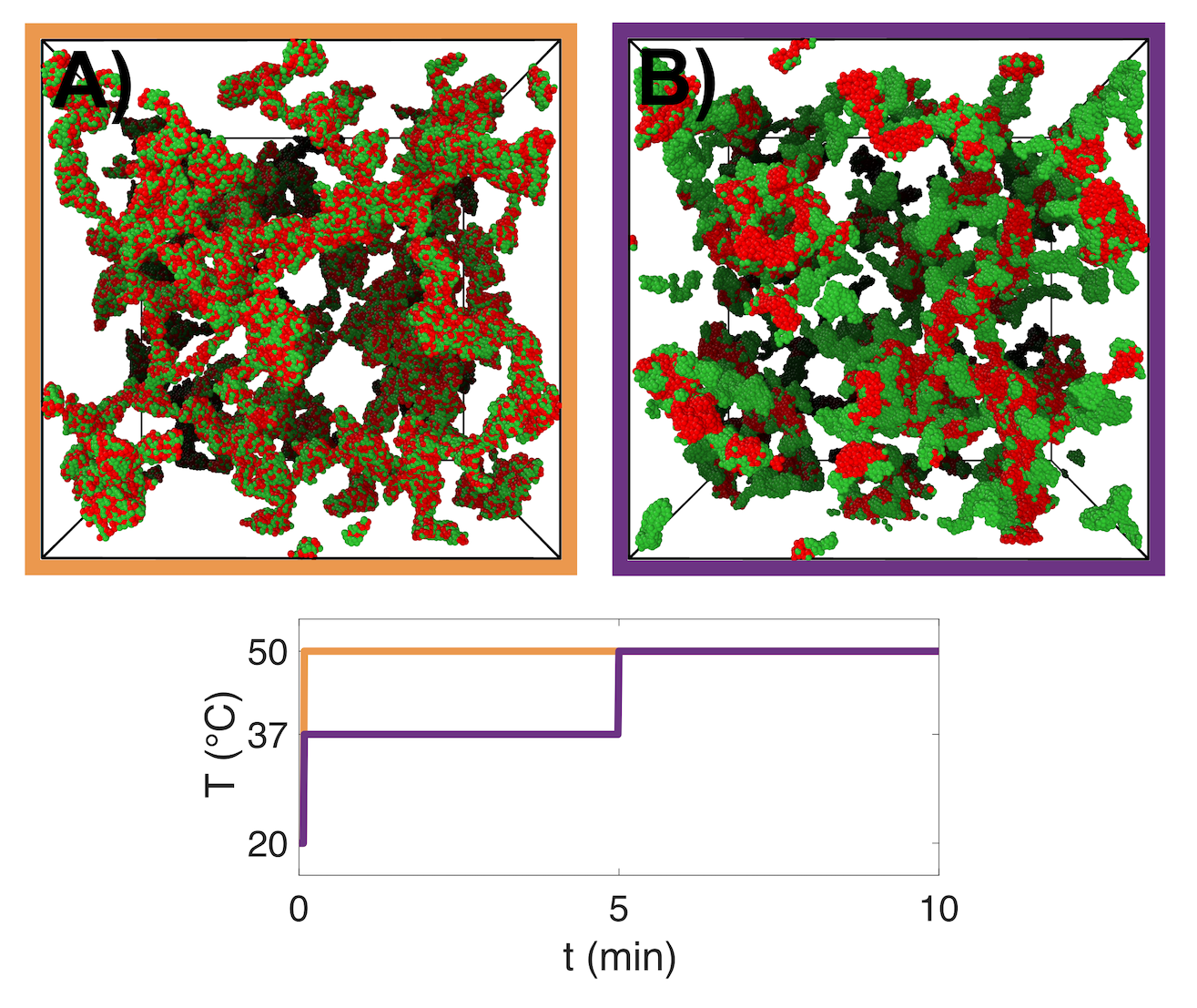}
\caption{(A): The final configuration of a single step quench simulation (orange temperature profile in bottom panel). (B): The corresponding final structure obtained through a two-step temperature quench (purple temperature profile).}
\label{RES-sim1st2}
\end{figure}

In order to further elucidate the self-assembly behavior observed in confocal microscopy, we performed Brownian dynamics simulations of binary microgel mixtures using a temperature-dependent pair potential. The effective potential depends three-fold on the temperature $T$: (\textit{i}) the particle diameter $\sigma$ decreases with $T$ due to the microgel collapse, (\textit{ii}) microgel stiffness is increased due to the increase in monomer density upon collapse,\cite{Richtering1999} and (\textit{iii}) the van der Waals attraction strength increases with $T$ as the monomer density of the particle increases. The total pairwise interaction $U_{\text{tot}}$ between microgel particles is described as the sum of a soft Hertzian repulsion $U_H$\cite{Schurtenberger2013,Lifshitz1986} and a van der Waals attraction $U_{\text{vdW}}$, and a detailed description of the model is found in the Methods section.
We first implemented a simulation which was allowed to equilibrate at 20~$^\circ$C, and thereafter was quenched instantaneously to 50~$^\circ$C. This quench causes all particles to exert the same attractive force at the same time. The final structure (seen in Fig.~\ref{RES-sim1st2}A) does not discern between the particle types, which is consistent with the behavior observed in CLSM for the fast quench system (Fig.~\ref{RES-confocalpic}C). Performing another simulation with an intermediate equilibration step at 37~$^\circ$C (Fig.~\ref{RES-sim1st2}B), one first obtains a single-species pN particle gel while pM microgels remain dispersed. Upon their collapse, pM particles turn fully attractive, and the strength of the attraction causes a diffusion-controlled cluster-cluster aggregation mechanism that yields growing pM clusters. These clusters will then bind to the pN gel strands upon encountering them (Fig.~\ref{RES-sim1st2}B). Comparing Figs.~\ref{RES-confocalpic}A and \ref{RES-sim1st2}B, it is clear that the employed temperature profile needs further refinement in order to reproduce the experimentally obtained structure: the monolayer of pM particles that is clearly seen decorating the pN gel in CLSM is absent in either simulation.

\begin{figure}
\centering
\includegraphics{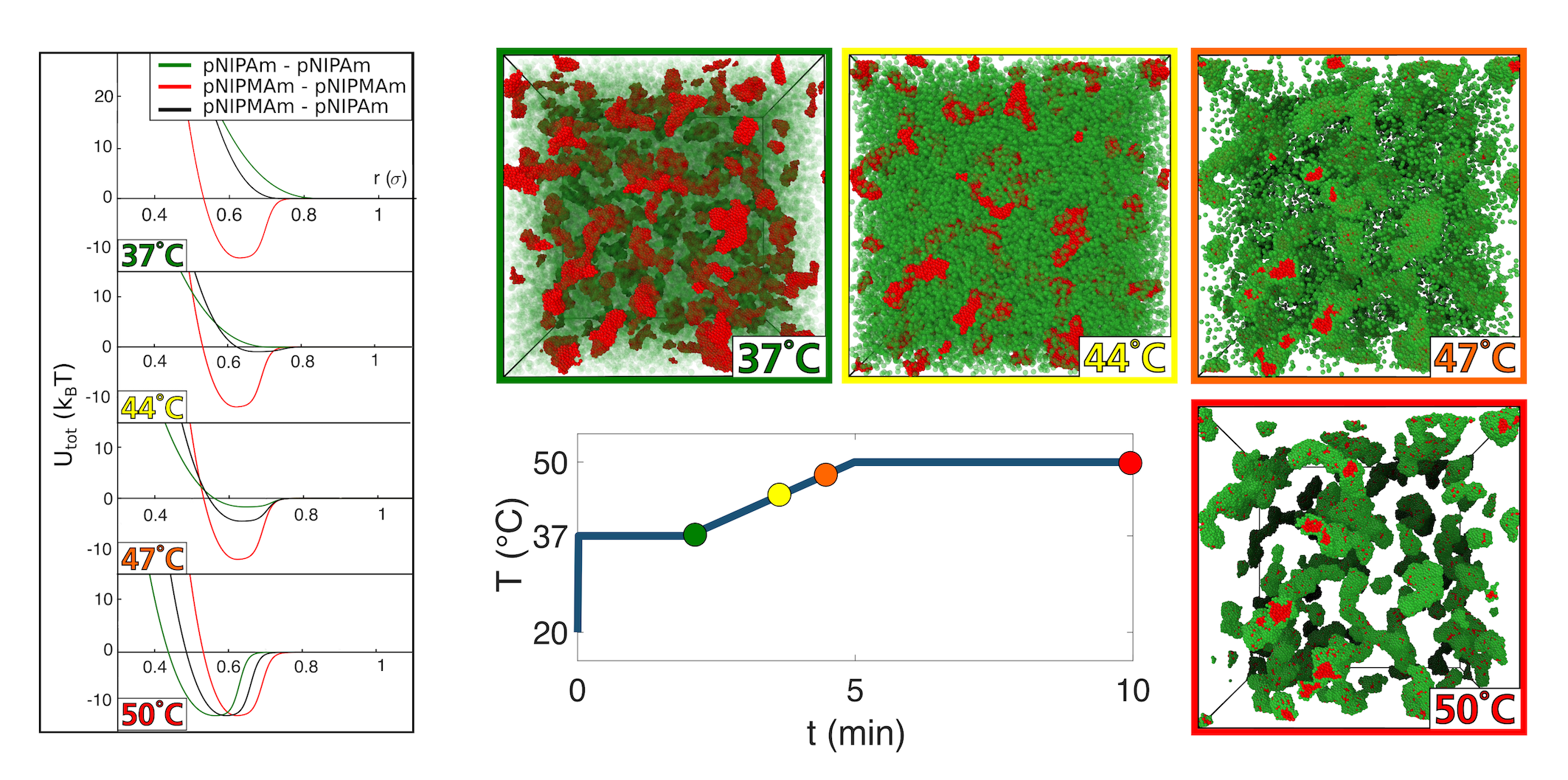}
\caption{Simulation snapshots taken at different times during a temperature ramp simulation, as indicated in the bottom panel. Above each snapshot, the three (pN-pN, pM-pM and pN-pM) instantaneous interaction potentials are shown. The size and transparency of the particles have been coupled to the swelling curves, with swollen particles being larger and more transparent.}
\label{RES-tempramp}
\end{figure}

We thus performed simulations using a slow temperature ramp where the $T$-dependent potential is gradually switched, comparable to the temperature ramp of Fig.~\ref{RES-confocalpic}A (see Fig.~\ref{RES-tempramp}). It is clear that slowly ramping through intermediate temperatures forces an adsorption of the pM particles onto the already present pN gel. The three effective pair potentials, shown in Fig.~\ref{RES-tempramp} $U(\text{pN-pN})$, $U(\text{pM-pM})$ and $U(\text{pN-pM})$, illustrate that, initially, pM particles preferentially bind to the pN gel network. Given enough time in a sufficiently slow ramp, they will thus prefer to form a monolayer shell around the existing network structure, leading to a core-shell structure that closely resembles the observations made by CSLM. Videos of these simulations can be found in the Supporting Information.$^\dag$\\

\subsection*{Rheology}
In order to compare the mechanical properties of the different gel structures, time-dependent oscillatory rheology was performed in the linear viscoelastic regime. For all single particle samples we used $\phi_{50} = 0.025$, and $\phi_{50}(\text{pN}) = \phi_{50}(\text{pM}) = 0.025$ for binary systems.
\begin{figure}
\centering
\includegraphics{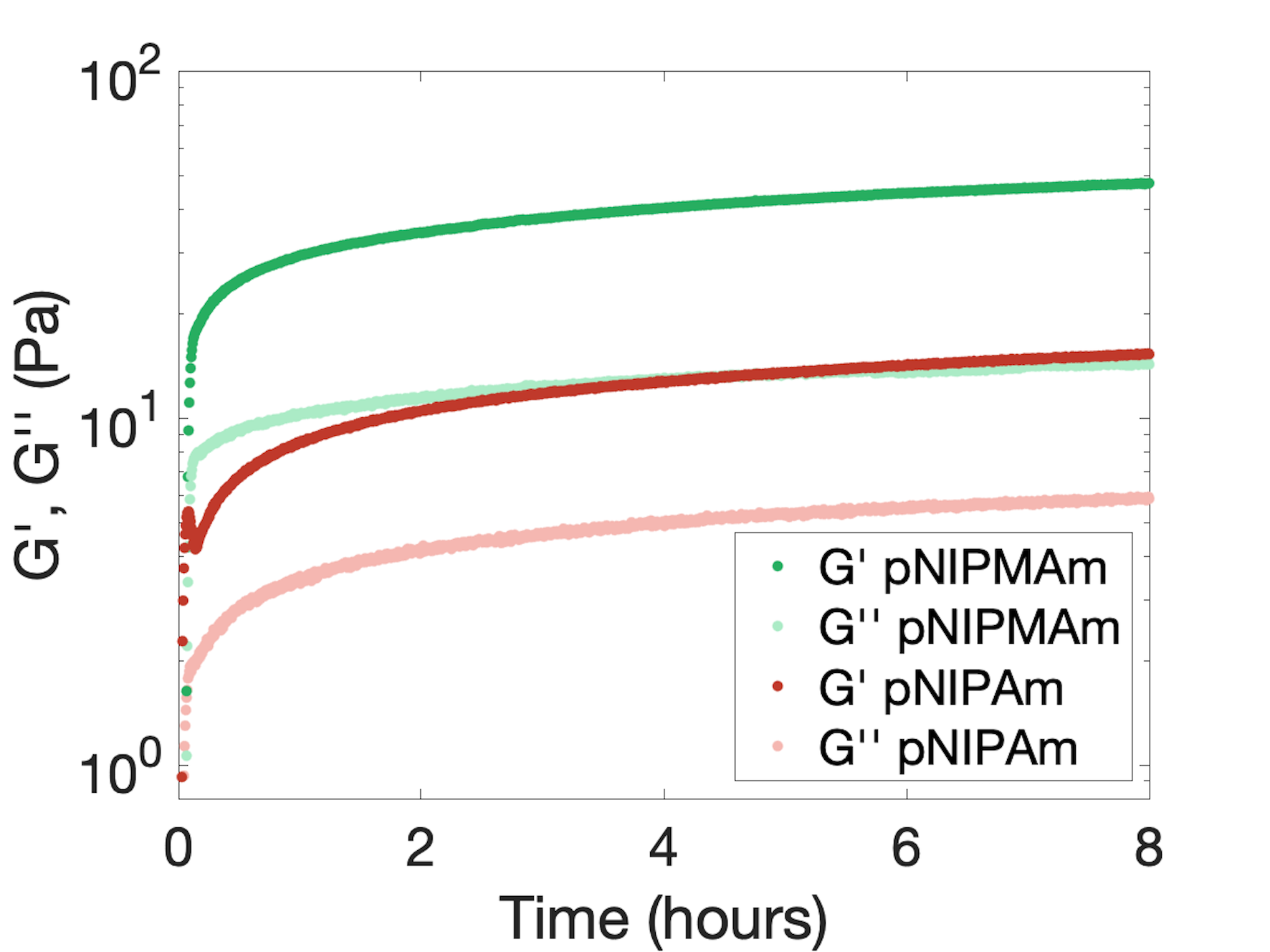}
\caption{Oscillatory rheological response at 1 Hz and 1 \% strain of single particle gels formed from either pN or pM microgel dispersions at $\phi_{50} = 0.025$, well inside the linear viscoelastic regime. The pM gel shows significantly higher moduli.}
\label{RES-pnippnipmcomp}
\end{figure}

Comparing the responses from pN and pM single particle gels, shown in Fig.~\ref{RES-pnippnipmcomp}, their elastic and loss moduli $G'$ and $G''$ both increase sharply upon crossing $T_{\text{VPT}}$ because of the subsequent gel formation, in accordance with the work by Liao \textit{et al.}\cite{Zhu2011} The magnitude of the pM response is significantly higher than that of its pN counterpart, likely caused by pM particles exerting stronger attractions compared to pN particles.\cite{Russel1991} Appel \textit{et al.} showed that elastic moduli in these types of gel scale with temperature above $T_{VPT}$,\cite{Sprakel2016} which might cause one to expect a lower $G'$ modulus for pM gels, since $[\text{50}~^\circ\text{C} -T_{VPT}(\text{pN})] > [\text{50}~^\circ\text{C} -T_{VPT}(\text{pM})]$. As we see the opposite, and given the very similar particle densities in the collapsed state and the similar molecular weights of either monomer resulting in highly similar Hamaker constants and thus van der Waals attractions, this effect has to be caused by specific differences between the two monomers, resulting in a stronger attraction between individual microgels. We believe that the basis of the additional attraction is most likely an attractive interaction between exposed methylated hydrophobic groups,\cite{Bureau2010,Christenson2001} where pM experiences this effect stronger due to the fact that it contains one additional methyl group per monomer.
\begin{figure}
\centering
\includegraphics{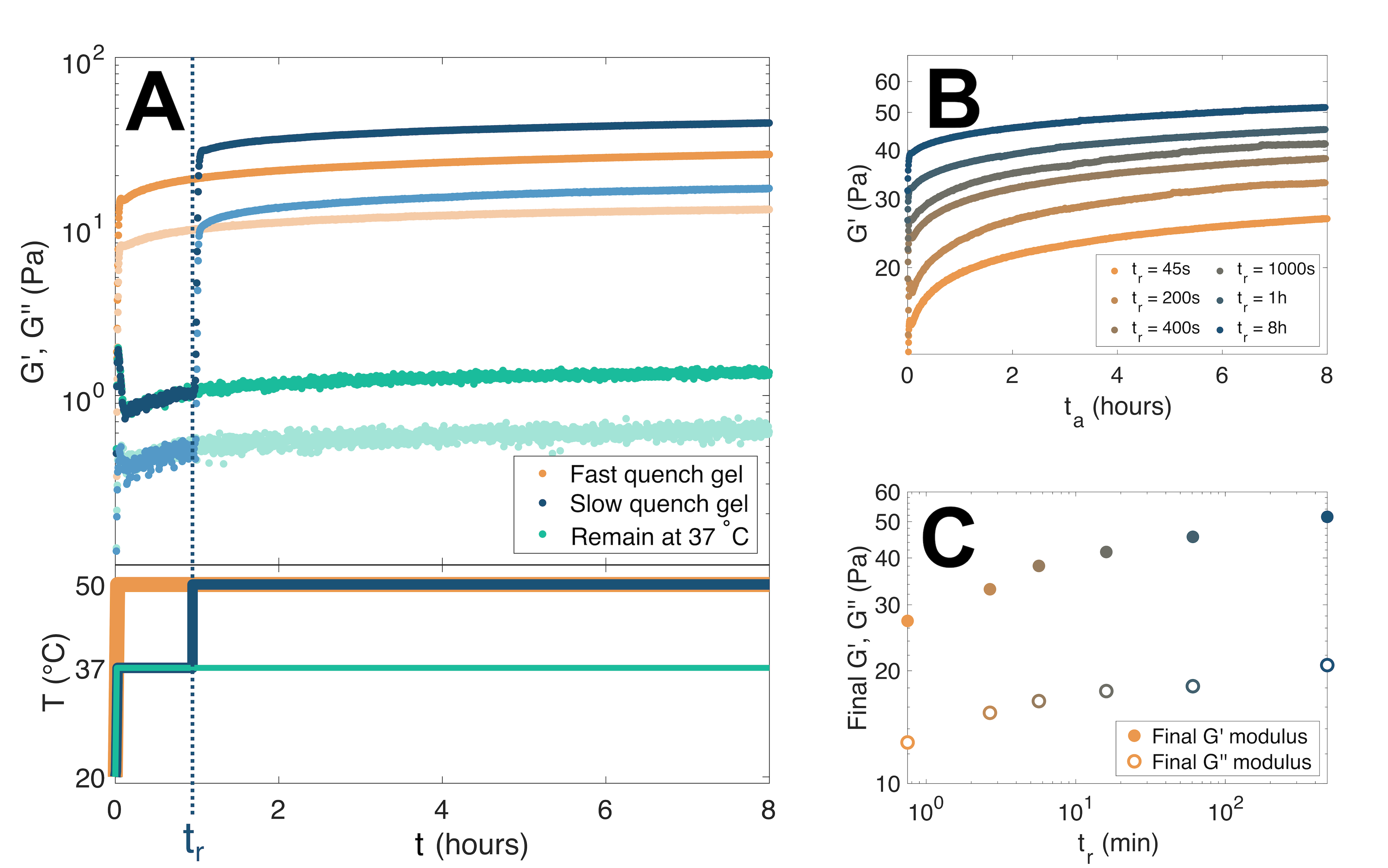}
\caption{(A) Oscillatory rheological response of gelled binary microgel dispersions of $\phi_{50}(\text{pN}) = \phi_{50}(\text{pM}) = 0.025$. Responses to three different temperature profiles are depicted: the darker symbols show elastic moduli, the lighter symbols show loss moduli. Temperature profiles have been added to the bottom of the graph. Note that slow quench gels yield a stronger response. (B) Elastic response of bigels with different rest times at intermediate temperatures $t_r$ as a function of aging time $t_a$, where $t_a = t - t_r$ is defined as the time the bigel is allowed to age at the final temperature of 50 $^\circ$C. The elastic response increases with the rest time. (C) $G'$ and $G''$ moduli of several gels after  $t_a = 8$ hours at 50 $^\circ$C, as a function of $t_r$. Closed symbols show elastic moduli and open symbols show loss moduli.}
\label{RES-bigel}
\end{figure}

A comparison between oscillatory rheology of slow and fast quench gels can be found in Fig.~\ref{RES-bigel}A. A slow quench was achieved by first heating the sample from 20~$^\circ$C to 37~$^\circ$C, maintaining it at 37~$^\circ$C for 60 minutes, followed by further heating to 50~$^\circ$C. The fast quench was achieved by a sudden flow of preheated thermostat water at 50~$^\circ$C, through which the sample reached 50~$^\circ$C in approximately 2 minutes. Comparing the responses shows that the slow quench core-shell network yields higher moduli than the more branched fast quench network. Additionally, as depicted in Fig.~\ref{RES-bigel}B, the rest time $t_r$, defined as the time where the sample is kept at an intermediate temperature between the two $T_{VPT}$ values of 32~$^\circ$C and 45~$^\circ$C (see also lower panel of Fig.~\ref{RES-bigel}A), strongly affects the elastic and loss moduli of the final gel (where aging time $t_a \equiv t - t_r$); the mechanical properties of the bigel scale with the resting time over which the pN scaffold gel is allowed to age before it is subjected to the final temperature quench. This is further emphasized in Fig.~\ref{RES-bigel}C, where the elastic and loss moduli at the same aging time $t_a = 8$ hours are plotted against $t_r$. All structural properties of the scaffold gel, such as strand thickness or gel age \textit{at time $t_r$}, thus ultimately determine the final mechanical properties of the bigel.\\
Forming a pure pN gel creates an immediate short-term overshoot of elastic and loss moduli during the initial gel formation, before quickly dissipating and returning to an aging baseline, observed in Figs.~\ref{RES-pnippnipmcomp} and \ref{RES-bigel}A. We believe that this effect is caused by the temperature-dependent deswelling behavior specific to these particles which can be found in the Supporting Information.$^\dag$ We hypothesize that this effect is caused by the further collapse of particles after initial gelation slightly above $T_{VPT}$.  At this stage, an initial network has formed and a decrease in particle size will strain this network, causing rearrangements that release internal gel strain, lowering elastic and loss moduli. A possible reason for why this is not observed for pM gels is that stronger pM-pM interactions could inhibit rearrangement.

\begin{figure}
\centering
\includegraphics{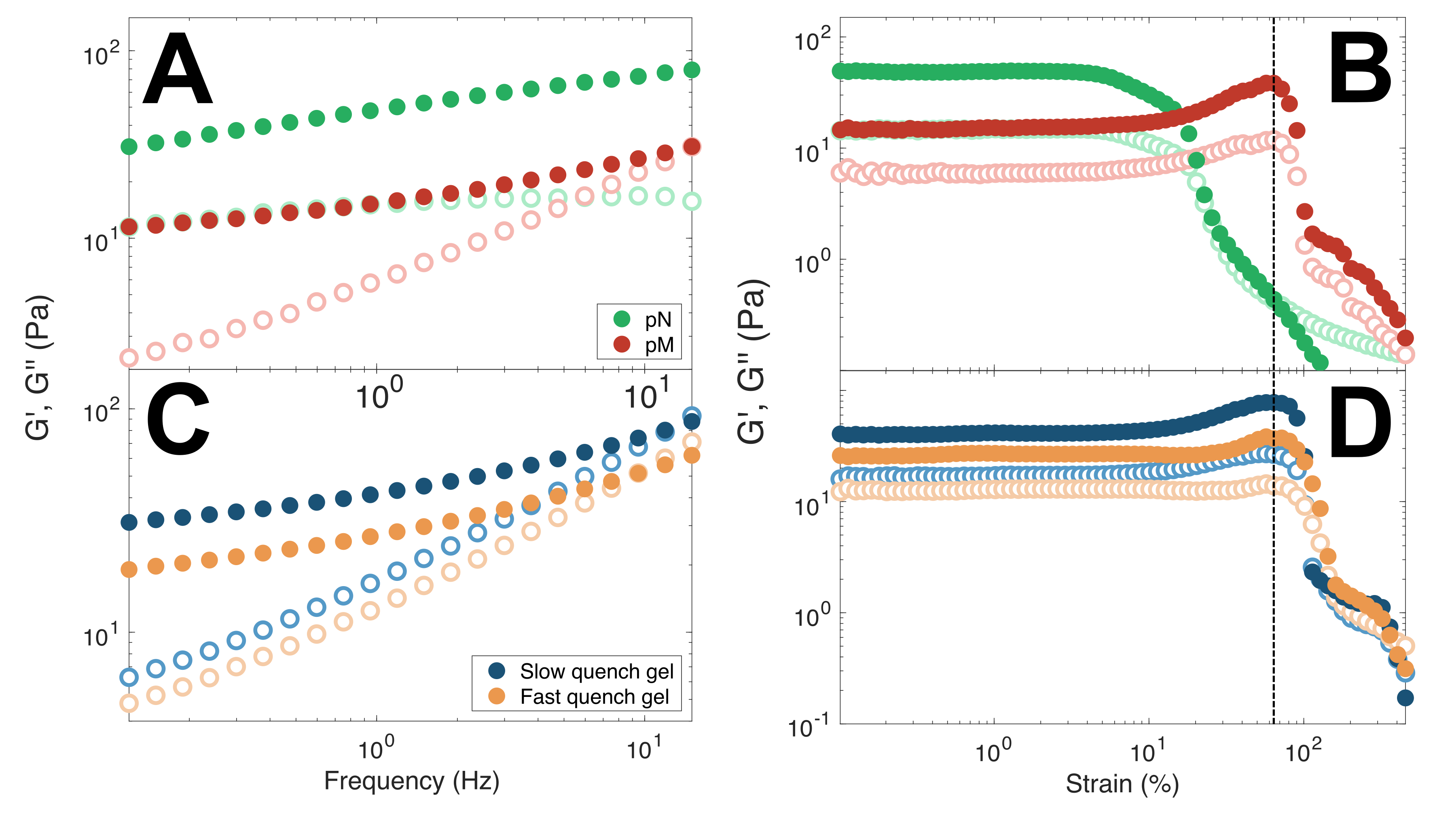}
\caption{(A) Frequency dependent oscillatory responses of gelled single microgel dispersions of $\phi_{50}(\text{pN}) = \phi_{50}(\text{pM}) = 0.025$. pM shows a different frequency dependence and a stronger behavior than its pN counterpart. (B) Yielding behavior of the same gels; pM is stronger but more brittle. (C) The frequency dependence of a slow and a fast quench gel; both systems have a similar frequency dependence as the scaffold pN gel, only the magnitude is different. (D) Strain sweeps of both bigels, which both fracture in a similar fashion to the scaffold pN gel. The black dashed line allows comparison of the yield strain. Closed circles denote $G'$ and open, brighter circles denote $G''$.}
\label{RES-bigelstrain}
\end{figure}

We furthermore compare the frequency dependence and yielding behavior of both bigel types with the single component gels. The frequency dependence of single component pN and pM particle gels, as seen in Fig.~\ref{RES-bigelstrain}A, reflects the stronger pM-pM attraction in both its difference in magnitude and in the fact that the pM response shows more solidlike behavior.\cite{Weitz2000,Egelhaaf2009} In Fig.~\ref{RES-bigelstrain}B, it can be seen that the pM particle gel is stronger initially but yields at lower strains, in correspondence with previous work on the effect of attraction strength on the yielding of colloidal gels.\cite{Zukoski1997,Zukoski2003,Petekidis2011,Osuji2013,Petekidis2014} If we compare these observations to the bigel systems, we observe that either bigel shows similar frequency dependence (Fig.~\ref{RES-bigelstrain}C) to the scaffold pN particle gel, rather than to its pM counterpart. Furthermore, in Fig.~\ref{RES-bigelstrain}D it can be seen that the yielding behavior of the bigel, while different in magnitude, has highly similar yielding points to its scaffold gel. We therefore conclude that the incorporation or deposition of pM onto a pN gel enhances the strength of the gel, while preserving the other characteristics of the scaffold gel.

\begin{figure}
\centering
\includegraphics{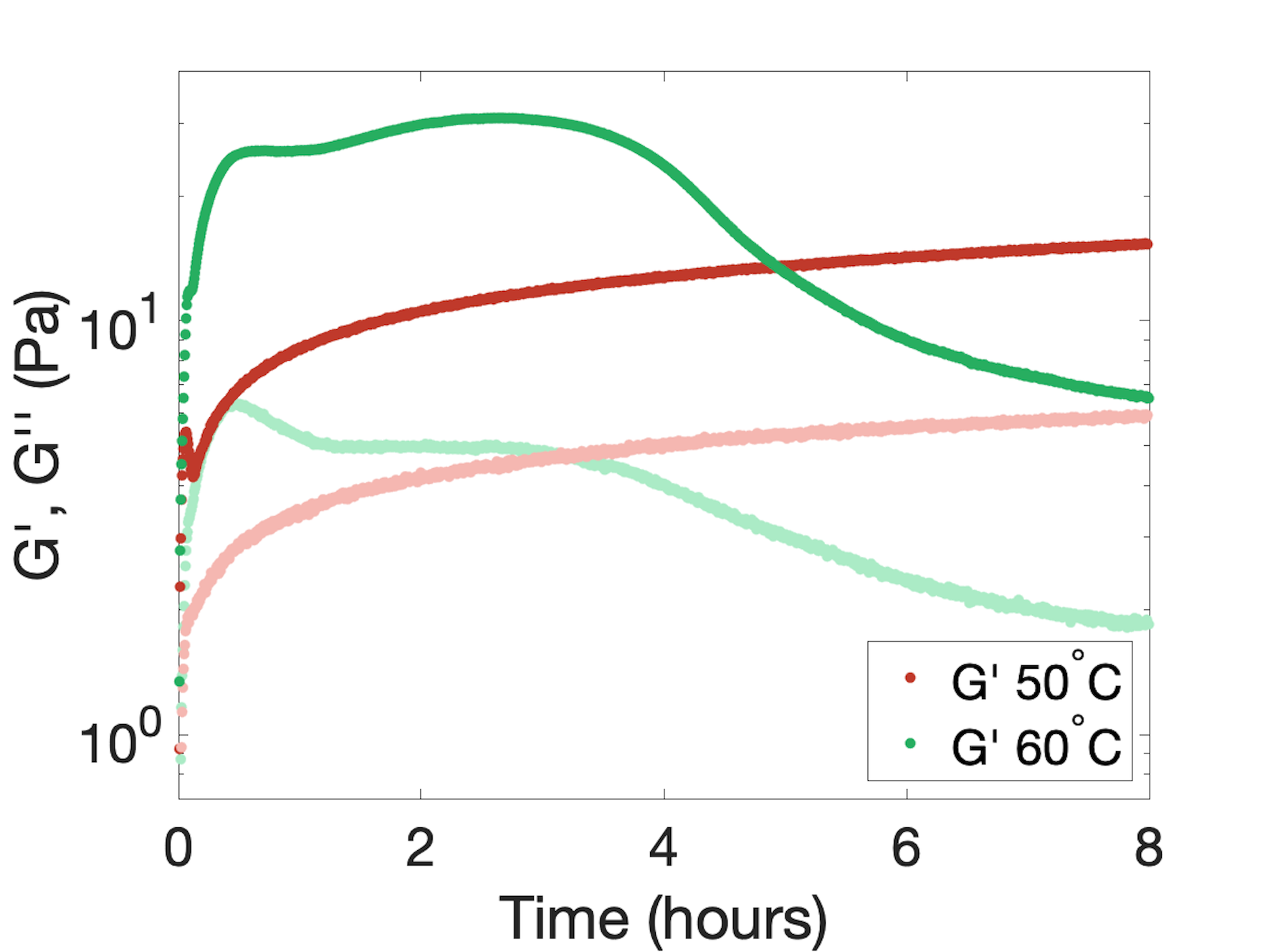}
\caption{Oscillatory rheological response of pN single particle gels at 50~$^\circ$C and at 60~$^\circ$C. When significantly above a certain failure temperature, a loss in moduli is observed due to gel failure.}
\label{RES-pnipfailure}
\end{figure}

A failure of pN gels is observed when the dispersion is heated significantly above $T_{\text{VPT}}$, seen in Fig.~\ref{RES-pnipfailure}, ending with optically visible sedimentation. We hypothesize that our gel failure arises due to homogeneous shrinking recently observed and discussed by Bischofberger \textit{et al.},\cite{Trappe2015} and that the application of shear breaks up destabilized gel bonds, ultimately yielding failure and sedimentation. Bischofberger \textit{et al.} discuss macroscopic shrinking of pNIPAm colloidal gel structures above $T_{\text{VPT}}$, while retaining the original shape. A hypothesis given for this behavior is the uniform local shrinking of gel scaffold elements, thereby expelling water from the interior of the macroscopic gel and a subsequent shrinking of the overall space occupied by the gel. A further collapse of particles when significantly above $T_{\text{VPT}}$ has not been observed in other studies\cite{Richtering1999, Wang1998} and thus cannot account for this effect. This shrinking behavior is distinctly different for the stronger pM gels: after probing a pure pM particle system at 70~$^\circ$C for 12 hours no decrease in elastic or loss moduli was observed.

\begin{figure}
\centering
\includegraphics{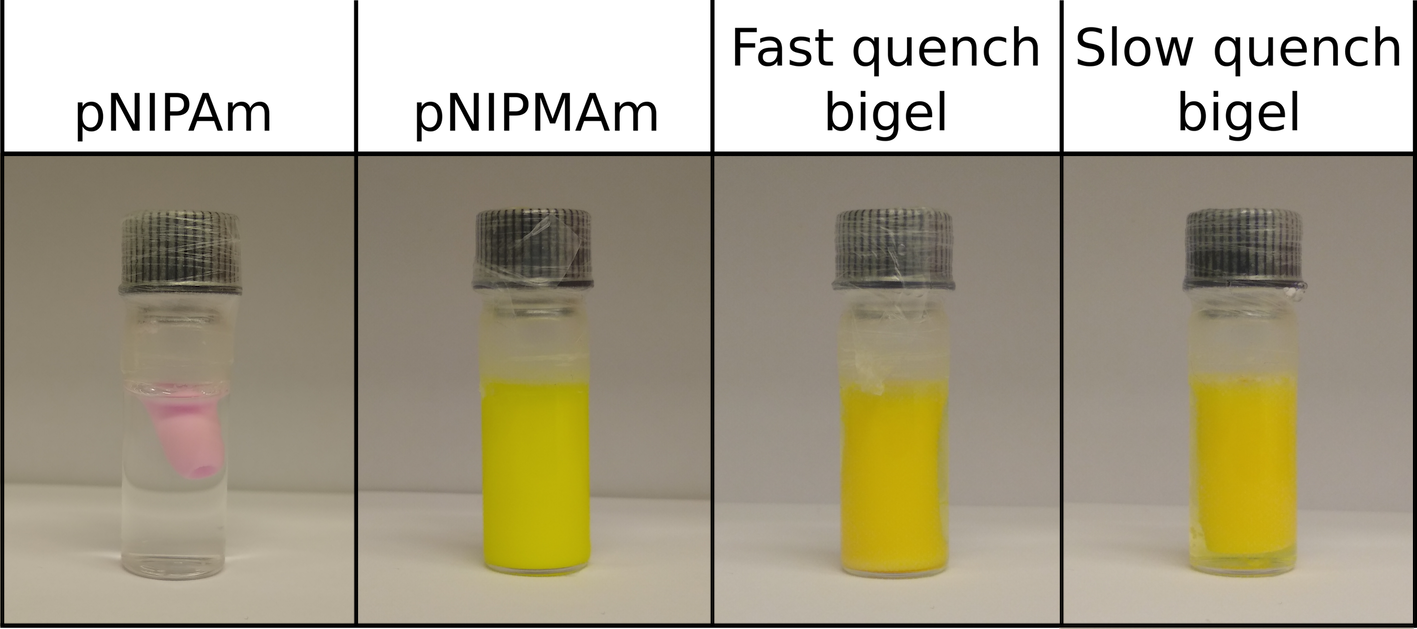}
\caption{Pictures of macroscopic pN and pM particle gels after heating at 60~$^\circ$C for 8 hours. The pure pN gel contracts significantly, while the corresponding pM gel does not visibly contract. In both binary gels, the pM particles reinforce the structure to reduce contraction over longer periods of time.}
\label{RES-Macrogel}
\end{figure}

Neither the pure pM gels nor the slow and fast quench bigels experience the failure and sedimentation occurring in single pN gels (Fig. ~\ref{RES-pnipfailure}). To further connect this to the macroscopic shrinking behavior, the macroscopic gel contraction over time was studied through similar experiments as described by Bischofberger \textit{et al.}:\cite{Trappe2015} two binary suspensions of $\phi_{50}(\text{pN}) = \phi_{50}(\text{pN}) = 0.025$, a pure pN suspension of $\phi_{50} = 0.05$ and a pure pM suspension of $\phi_{50} = 0.05$ were prepared and put in a water bath at 60~$^\circ$C. The contraction of the macroscopic gel was followed over the course of 8 hours (see Fig.~\ref{RES-Macrogel}). During this period, pN gels lost approximately 80 \% of their volume, pM gels showed no visible contraction, while the binary gels both lost approximately 10 \% of their volume. It is clear that incorporation of pM in a pN gel strongly affects its time-dependent structural properties and protects it from degrading aging processes, most likely caused by the stronger pM particle attraction.

\section*{Conclusions and Perspectives}
In the present work, we have presented a binary colloidal gel whose structural and mechanical properties can be externally controlled through different heating profiles. We have shown that the frequency- and strain dependent behavior of bigel mechanical properties conforms to the properties of the scaffold gel, but the strength of the gel is enhanced by secondary particle deposition. Furthermore, the final mechanical properties are affected by the structural properties of the initial scaffold structure at the moment of secondary particle deposition: by allowing aging to affect the scaffold structure before secondary particle deposition, we can control the mechanical properties of the resulting bigel. We have furthermore shown confocal laser scanning microscopy images showing distinctly different structures depending on the heating profile used. For the slow quenches a bigel conforms to a scaffold structure, while for fast quenches a thinner and more branched network forms. Brownian dynamics simulations were used to mimic the temperature ramp-dependent structure formation and demonstrate the crucial role played by the temperature profile in structure formation.

Binary gels made from thermoresponsive microgels allow control over a number of interesting features. Due to the fact that we can carefully tune the effective pair potentials between the different components, we have exceptional control over the two structural length scales that define mechanical properties of colloidal gels. Choosing the concentration of the first particle with the lower transition temperature and quenching the temperature to a value above its $T_{\text{VPT}}$, but below the $T_{\text{VPT}}$ of the second species, will define the mesh size or correlation length of the resulting gel. A subsequent slow increase of the temperature will then destabilize the second particle species and will result in the formation of a shell around the existing network and a corresponding increase of the thickness of the network strands, while preserving the overall structure. Here it is also important to point out that this sequential adsorption step could easily be repeated with other microgel species having an even higher $T_{\text{VPT}}$,\cite{Kaoru2005,Hu2010} thus allowing for the formation of several shells and a corresponding reinforcement of the overall mechanical properties. Together with the fact that the process is fully reversible and that microgels offer almost unlimited possibilities for added functionalization through the incorporation of specific comonomers or nanoparticles with specific optical (\textit{e.g.} plasmonic), catalytic or magnetic properties,\cite{Hu2010,Hoare2004,Zhou2015,Karg2009,Das2007,Liu2012,Backes2015} these systems provide highly interesting routes to materials with vastly improved properties.

\section*{Methods}
\subsection*{Microgel particles}
Microgels used in this work were prepared using a method by radical polymerization of monomer in the presence of 5 mol\% cross-linking agent, as described previously.\cite{Mansson2015} Several important properties of samples described in this work are summarized in Table~\ref{Methods-parttable}. The monomers used were N-isopropylacrylamide and N-isopropylmethacrylamide (97\%), the initiator was potassium persulfate (99\%) and dyes used were covalently binding fluorescein O-methacrylate (97\%) or methacryloxyethyl thiocarbamoyl rhodamine B (97\%). All particles were prepared in the presence of 5 mol\% N,N'-methylene-bis-acrylamide (99\%) cross-linker. All chemicals were purchased from Sigma Aldrich. Hydrodynamic radii as function of temperature were determined with dynamic light scattering using a modulated 3D cross-correlation instrument (LS instruments) with a 660 nm diode pumped laser. It should be noted that hydrodynamic radii were obtained at $10^{-3}$M KCl in order to prevent aggregation for temperatures above $T_{\text{VPT}}$, leading to an overestimation of radii compared to fully screened conditions. All particles are monodisperse in size and crystallize at high volume fractions. The number density of stock solutions was thus determined from fully crystalline dispersions at 20~$^\circ$C by counting the number of particles $n_p$ in single crystals using CLSM imaging. The latter quantity was then used to estimate the effective volume fraction $\phi_{50}$ considering the hydrodynamic diameter $\sigma_{H,50}$ at 50~$^\circ$C  of the particles following:
\begin{equation}
\phi_{50} = \frac{N_p \pi \sigma_{H,50}^3}{6}
\end{equation}

\subsection*{Experiment and simulation}
All samples described in this work contain 10$^{-1}$M KCl, in order to effectively screen Coulomb interactions. All described binary samples have $\phi_{50}(\text{M}_\text{pN}) = \phi_{50}(\text{M}_\text{pM}) = 0.025$, and for single particle samples either $\phi_{50} = 0.025$ or $\phi_{50} = 0.05$.

\subsubsection*{Confocal Laser Scanning Microscopy}
Microgel dispersions for confocal laser scanning microscopy (CLSM) were held in a custom-made capillary cell, prepared from glass slides and UV-glue, that allow for water- and air-tight conditions for approximately one month. Cells were mounted on an inverted CLSM (Leica TCS SP5 tandem scanner), and imaged using a 100x oil immersion objective. The microscope was mounted in an enclosure that allows for temperature control with a 0.2~$^\circ$C maximum variance using thermostated air circulation. Samples can be heated by contact with a preheated metal block, or by contact with a double-walled glass container which is connected to a thermostated water bath of identical make as the rheometer water bath. Final images had their fluorescent background removed using the "Subtract background" option in ImageJ.\cite{Cardona2012}

\subsubsection*{Particle interactions}
The total pairwise interaction $U_{\text{tot}}$ between microgel particles of the same species is described as the sum of a soft Hertzian repulsion $U_H$ and a van der Waals attraction $U_{\text{vdW}}$, where the former is given by

\begin{gather}
U_H(r;T) = 
\begin{cases}
\epsilon_H(T) \left(1 - \frac{r}{\sigma(T)}\right)^{\frac{5}{2}} & \text{if } r \leq \sigma \\
0 & \text{if } r > \sigma
\end{cases},
\label{eq-Hertz}
\end{gather}
where $r$ is the particle center-to-center distance, and
\begin{equation}
\epsilon_H(T) = \frac{2E(T)}{15K} \frac{1}{1-\nu(T)^2} \sigma(T)^3.
\label{eq-Lifsh}
\end{equation}
In Eq.~\eqref{eq-Lifsh}, $E(T)$ is the temperature-dependent Young's modulus. We use the Young's modulus determination for pNIPAm microgel particles from Fernandes \textit{et al.}\cite{Hellweg2010} and fit a hyperbolic tangent function to their experimental data, according to
\begin{equation}
E(T) = C_1 \text{tanh}\left(\frac{T - T_c}{C_3}\right) + C_2,
\label{eq-tanhE}
\end{equation}
where $C_1 = 2.6\cdot10^5$ Pa, $C_2 = 3.3\cdot10^5$ Pa, $T_c = 310$ K and $C_3 = 3$ K for pN particles. For pM we used the same constants except $T_c = 323$ K. Furthermore, the Poisson ratio $\nu$ is described by the same hyperbolic tangent as in Eq.~\eqref{eq-tanhE} but with $C_1 = 0.125$, $C_2 = 0.375$, $T_c = 310$ K and $C_3 = 3$ K, yielding $\nu(20$~$^\circ C) = 0.25$ and $\nu(50$~$^\circ C) = 0.5$.\cite{Schurtenberger2017} The empirical scaling factor $K$ in Eq.~\eqref{eq-Lifsh} was adjusted so to yield the value 496 $k_BT$ at 20~$^\circ$C as obtained by Paloli \textit{et al.}\cite{Schurtenberger2013} Finally, the particle diameter $\sigma(T)$ was determined from the hydrodynamic radii of pN and pM particles as determined by dynamic light scattering.

The van der Waals attraction $U_{\text{vdW}}$ is furthermore given by a short-range, rounded square well potential, described by:
\begin{equation}
U_{\text{vdW}}(r;T) = \frac{\epsilon_{\text{vdW}}}{2}\tanh\left(\frac{r - 1.1\sigma}{k}\right) + \frac{\epsilon_{\text{vdW}}}{2};
\label{eq-vdW}
\end{equation}
\begin{equation}
\epsilon_{\text{vdW}}(T) = \epsilon_{\text{max}} \left(\frac{\sigma(20~^\circ\text{C}) - \sigma(T)}{\sigma(20~^\circ\text{C}) - \sigma(50~^\circ\text{C})}\right)^{6},
\label{eq-vdWeps}
\end{equation}
where the $\sigma^{6}$ dependence of the attraction strength $\epsilon_{\text{vdW}}$ comes from the fact that the van der Waals interactions depend on $\rho^2$, where $\rho$ is the monomer density of the particles. The maximum attraction $\epsilon_{\text{max}}$ was set to 12 $k_BT$ in accordance with measurements on comparable particles.\cite{Ballauff2011, Zaccone2013} In order to calculate the interaction between two particles of different species, we used the Lorentz-Berthelot mixing rules $\epsilon_{\text{AB}} = \sqrt{\epsilon_\text{A}\epsilon_\text{B}}$ and $\sigma_{\text{AB}} = \frac{\sigma_\text{A} + \sigma_\text{{B}}}{2}$.

\subsubsection*{Brownian dynamics simulations}
All simulations were performed using the open-source LAMMPS molecular dynamics package,\cite{Plimpton1995} using the overdamped Langevin dynamics for the position $\textbf{r}_i$ of particle $i$:

\begin{equation}
\frac{d\textbf{r}_i}{dt} = -\frac{\textbf{F} (\{ \textbf{r}_j \}) D}{k_BT} + \sqrt{2D}\textbf{R}(t),
\end{equation}
where $D$ is the single particle diffusivity and $\textbf{R}(t)$ is a unit-variance random noise, $\delta$-correlated in space and time. All simulations were run with $N = 100,000$ particles at a volume fraction of $\phi_{50} = 0.05$ at 50~$^\circ$C in order to mimic experiment.\\
Periodic boundary conditions were applied in all three dimensions. In order to map the time units of our simulations to experimental time scales, we define the Brownian time $\tau_B$ as:
\begin{equation}
\tau_B \equiv \frac{\sigma^2(T)}{D(T)} = \frac{3\pi\mu(T) \sigma^3(T)}{k_BT}
\end{equation} 
with $\mu$ the dynamic viscosity, $k_B$ the Boltzmann constant and $T$ the temperature. In real units, $\tau_B$ ranges from 0.56 s at 20~$^\circ$C to 0.051 s at 50~$^\circ$C. Visualizations were obtained with the Open Visualization tool OVITO.\cite{Ovito}
 
\subsubsection*{Rheology}
Rheology was performed using a TA Instruments ARES strain-controlled rheometer using a couette geometry with 10 mL sample volume, fitted with an external temperature bath. Temperature adjustments for single particle gels were achieved by simple ramping of 5~$^\circ$C~min$^{-1}$. Temperature adjustments for slow quench bigels were obtained ramping to 37~$^\circ$C with 5~$^\circ$C~min$^{-1}$, resting for a given time $t_r$, and further heating to 50~$^\circ$C with the same ramping speed. Temperature adjustments for fast quench bigels were achieved by preheating the thermostat to 50~$^\circ$C while clamping off the connection to the sample chamber, and a subsequent and sudden release of the clamp. Samples were measured using 1.0\% strain at a frequency of 1 Hz for time dependent measurements. Frequency-dependent measurements were performed after the time-dependent measurements with $t_a = 8$ hours and at 1\% strain, and strain-dependent measurements were performed thereafter at a frequency of 1 Hz. A solvent trap was applied to minimize evaporation effects.

\section*{Acknowledgments}
We gratefully acknowledge financial support from the European Research Council (ERC-339678-COMPASS) and the Swedish Research Council (2014-4037). JS is financed by a grant from the Swedish Research Council (2015-05449). Maxime Bergman is acknowledged for helpful discussions regarding image analysis, Niels Boon is acknowledged for helpful discussions regarding the model, and Linda M{\aa}nsson is acknowledged for particle synthesis. Simulations were performed on resources provided by the Swedish National Infrastructure for Computing (SNIC) at LUNARC.

\section*{Supporting Information}
\textbf{Supporting Information Available:} Videos of simulations described in Section "Model and Simulation", showing the single step quench simulation (final structure in Fig.~\ref{RES-sim1st2}A), the two-step temperature quench simulation (final structure in Fig.~\ref{RES-sim1st2}B) and the temperature ramp simulation (corresponding to Fig.~\ref{RES-tempramp}). Swelling curves of pN and pM particles. This material is available free of charge \textit{via} the Internet at http://pubs.acs.org.

\footnotesize{
\bibliography{bib_rheo1}
}

\pagebreak

\end{document}